\documentstyle[12pt,aasms]{article}
\tighten          %single space
\newcommand{\beq}{\begin{equation}}
\newcommand{\eeq}{\end{equation}}
\newcommand{\beqa}{\begin{eqnarray}}
\newcommand{\eeqa}{\end{eqnarray}}

\newcommand{\dv}{$\Delta {\rm V} ({\rm TO} - {\rm HB})$ }

\newcommand{\teffa}{T_{\rm eff} }

\newcommand{\ea}{{\it et al. }}

\newcommand{\Li}{\hbox{$^7$Li}}
\newcommand{\be}{\hbox{$^9$Be}}
\newcommand{\li}{\hbox{$^7$Li~}}
\newcommand{\Lie}{{\cal L}_{\lower4pt\hbox{$\!\!\!\!\scriptstyle\xi$}}}
\newcommand{\feh}{[{\rm Fe}/{\rm H}]}
\newcommand{\nli}{\log {\rm N} ({\rm Li})}

\slugcomment{to appear in ApJ}

\begin{document}

\title{$^7$Li Abundances in Halo Stars: \\
Testing Stellar Evolution Models\\
and the Primordial \li Abundance }

\author{Brian Chaboyer\altaffilmark{1,2} and  P. Demarque\altaffilmark{1,3}}

\altaffiltext{1}{Department of Astronomy, and Center for Solar and Space
Research, Yale University, Box 208101, New Haven, CT 06520-8101}

\altaffiltext{2}{present address: CITA, 60 St. George St., U. of
Toronto, Toronto, Ontario, Canada M5S 1A7 Electronic Mail --
I:chaboyer@cita.utoronto.ca}

\altaffiltext{3}{Center for Theoretical Physics, Yale University\\
Electronic Mail --I:demarque@astro.yale.edu}

\begin{abstract}
A large number of stellar evolution models with $\feh = -2.3$ and
$-3.3$ have been calculated
in order to determine the primordial \li abundance and to
test current stellar evolution models by a comparison to
the extensive database of accurate Li abundances in extremely
metal poor halo stars observed by Thorburn (1994).
Standard models with grey atmospheres
do a very good job of fitting the  observed Li abundances in stars
hotter than $\sim 5600$ K.  They predict a primordial \li abundance
of $\nli = 2.24\pm 0.03$. Models which include
microscopic diffusion predict a downward curvature
in the \li destruction isochrones at hot temperatures which is not
present in the observations.  Thus,
the observations clearly rule out models which include uninhibited
microscopic diffusion of \li from the surface of the star.
Rotational mixing  inhibits the microscopic diffusion
and the  $\feh = -2.28$ stellar models which include both diffusion and
rotational mixing provide an excellent match to the mean trend in
$\teffa$ which is present in the observations. Both the plateau stars
and the heavily depleted cool stars are well fit by these models.
The rotational mixing leads to considerable \li depletion in these
models and the primordial \li abundance inferred from these models
is $\nli = 3.08\pm 0.1$.  However, the $\feh = -3.28$
isochrones reveal problems with the combined models.  These isochrones
predict a trend of decreasing $\nli$ with increasing $\teffa$ which
is not present in the observations. Possible causes for this
discrepancy are discussed.
\end{abstract}


\section{Introduction}
Standard big bang nucleosynthesis (BBN) has been remarkably
successfully  in predicting the abundance of the various light
elements ($^1$H, $^2$H, $^3$He, $^4$He, \Li).  In standard BBN,
consistency with the inferred primordial abundance of $^2$H, $^3$He
and $^4$He requires the primordial \li abundance  in the range
$1.9 < \nli < 2.3$ (Krauss \& Romanelli 1990;
Walker \ea 1991; Smith Kawano \& Malaney 1993).
In recent years, various
alternatives to the standard big bang have been proposed (cf. the
review by Malaney \& Mathews 1993).  Perhaps the
most important of these are the inhomogeneous theories, which assume
that density inhomogeneities exist prior to nucleosynthesis (Witten
1984).  In general, inhomogeneous BBN models produce far more \li than
standard BBN.  Thus, the primordial \li abundance provides an
important test of BBN. Observations of Li\footnote{In stars, Li may
exist in two different isotopes: $^6$Li and \Li.  It is extremely
difficult observationally to distinguish between $^6$Li and $^7$Li,
thus the observations typically determine the total Li present in a
star.  Smith, Lambert \& Nissen (1992) determined the abundance ratio
of $^6$Li/\li to be $0.05\pm 0.02$ in the halo star HD 84937.  It
appears that the total Li content in a star is dominated by \Li.
Thus, we will assume that the Li observations actually measure \Li.}
in  old, metal poor hot stars ($\feh \la -1.5$, $T_{\rm eff} > 5590$ K)
in our Galaxy have found a  nearly uniform abundance of $\nli \simeq
2.2$ observed in  hot stars (Spite \& Spite 1982, 1986; Rebolo,
Molaro \& Beckman 1988; Hobbs \& Thorburn 1991).  This has been viewed as
confirmation of standard BBN.

Unfortunately, this neat match between observation and theory has been
brought into question by observations of extremely metal-poor stars
([Fe/H] $ < -1.9$). Over 70 such stars have been observed (Thorburn
1994) and 3 of them show a significant \li depletion.  Thus, some
mechanism is destroying \li in at least some metal-poor stars.
This may be taken as a sign that a few stars undergo an very efficient
\li depletion process, such as a stellar merger, which does not affect
the large majority of plateau stars (Hobbs, Welty, \& Thorburn 1991;
Spite \ea 1993).
However, there is convincing evidence from  studies of young
cluster stars   that \li depletion occurs on the main sequence
in  metal rich stars ($\feh > -0.1$)  (Thorburn \ea 1993;
Soderblom \ea 1993; Chaboyer, Demarque \& Pinsonneault 1994a; 1994b,
hereafter CDP).   These
facts suggest that a careful study must be made of possible mixing
mechanisms in metal-poor stars in order to interpret the observed \li
abundances.  Calculations by
Pinsonneault, Deliyannis \& Demarque (1992) suggest that rotational
mixing may lead to a depletion
of \li by at least a  factor of ten, in {\it all\/} metal-poor stars.
Such a large change in the primordial \li abundance could lead to an
incompatibility with standard BBN.  This has far reaching consequences
regarding the early history of our universe.
However, the great uncertainty in the rotational mixing coefficients
prevented Pinsonneault \ea from making  definitive conclusions regarding
the primordial \li abundance.

In addition to being an important probe of primordial nucleosynthesis,
the abundance of \li in the halo stars serves as another test of
stellar evolution models.   Chaboyer (1993) examined the issue of
mixing in stellar radiative zones by comparing stellar evolution
models to \li abundances and rotation velocities in the Sun and young
cluster stars with $M \le 1.3~M_\odot$.  It was found that standard
models, and models which
included microscopic diffusion did not deplete enough \li on the main
sequence.  Stellar models which included rotational mixing as well as
diffusion were able to reproduce the observed \li depletion pattern.
These results are presented by CDP.
The halo stars provide a severe test for our models, for a nearly
uniform abundance of
\li is observed over a wide range in effective temperature
($5600 \la T_{\rm eff} \la 6450$) and metallicity, $-1.30\la \feh \la
-3.80$ (hereafter referred to as the `plateau').  Thus, any mixing
mechanism which leads to the destruction of
\li must do so in a way which is independent of the stars temperature
(mass) and metallicity.

The \li depletion patterns in  metal poor stellar models have been
studied extensively by Deliyannis \& Demarque (1991a,b) who included
microscopic diffusion in an approximate manner, and Pinsonneault \ea
(1992) who considered the effects of rotational mixing. However, the
interaction between diffusion and rotational mixing has
not been studied.   Proffitt \& Michaud (1991) and Chaboyer \ea (1992)
 studied  the effect of microscopic diffusion on the primordial
\li abundance in some detail.  Since these works were completed,
the opacities  have been updated (Iglesias \& Rogers 1991; Kurucz 1991),
new model atmospheres from Kurucz (1992) have become available and the
energy producing nuclear reaction rates have been updated (Bahcall \&
Pinsonneault 1992).   In addition, a large sample of
extremely metal poor stars have had their \li abundances determined by
Thorburn (1994), who detected \li in 78 stars with $\feh < -1.90$ and
found upper limits in 6 more stars.  Such a large sample of \li
abundances (which have been determined in a consistent manner) serve as an
ideal test of our stellar models and BBN.

In this paper, we examine the question of the
primordial \li abundance by constructing stellar models with
$Z = 10^{-4}$ and $10^{-5}$ ($\feh = -2.28$ and $-3.28$) and comparing
them to the observations. This paper complements CDP, where the stellar
models are compared to young clusters stars and the Sun, with
$-0.10 \le \feh \le +0.15$.  A future paper will present the
$^6$Li and Be depletion isochrones, compare them to
the observations, and comment on the constraints to BBN.

In \S\ref{sect6p1a} we present a short
discussion of the new observations made by Thorburn (1994).  The
construction of the stellar models and the importance of the
surface boundary conditions in determining the \li
depletion in cool stars is discussed in \S\ref{sectnew}
{}~We examine standard models in \S \ref{sect6p2},
models which include microscopic diffusion in \S \ref{sect6p3}
and models which include both rotational mixing and microscopic
diffusion in \S\ref{sect6p4}   ~A discussion of the primordial \li
abundance and an overview of the major conclusions of this paper
is presented in \S \ref{sect6p5}

\section{The Observations \label{sect6p1a}}
Thorburn (1994) determined accurate \li abundances for over 80 stars
which have
$\feh < -1.9$.  The typical error in $\nli$ is $\sim 0.1$ dex, which
is considerably smaller than previous work.  The large number and high
accuracy of these new observations reveal trends which were not
discernible with the previous data.  Three stars in the plateau star
temperature range ($T_{\rm eff} > 5590$ K) are highly depleted.  If
these three stars are ignored, then a two dimensional
fit to the plateau star
observations reveals a significant trend of increasing
$\nli$ with {\it both\/} $T_{\rm eff}$ and $\feh$.  Thorburn
determined slopes of $d\nli/d\feh = 0.134$ dex per dex and
$d\nli/dT_{\rm eff} = 3.4\times 10^{-4}$ dex per K.  In addition, the
data have a significant dispersion of $0.07$ dex in $\nli$ about these
straight line fits.

The trend with $\teffa$ is not too surprising, for previous data had
hinted that such a trend might exist.  However, the trend of
decreasing \li abundance with decreasing metallicity is unexpected.
This metallicity trend is present at a very high confidence level.
In determining trends with $\feh$ one must proceed with
caution, due to possible systematic errors in the temperature scale or
in the model atmospheres  which might correlate with metallicity.  If
such errors are present, they could easily yield the observed trend of
\li abundance with $\feh$.  If this trend is real, it has important
ramifications on our interpretation of the data.  Such a trend is
naturally explained by the hypothesis that \li production has occurred
by galactic cosmic ray fusion (Thorburn 1994).  As such, the initial
abundance of \li in the stars is not the primordial one, but has been
enhanced  due to the \li production. The observed dispersion in $\nli$
is the result of a dispersion of 2 Gyr in the halo $\feh$--age
relation.

An alternative explanation for  the increase in $\nli$ with
increasing $\feh$ is that the \li depletion in the models is larger
for models with lower metallicity.  However, as is shown later, all of
our models exhibit the opposite trend: the amount of \li depletion
decreases with decreasing metallicity.

Since the presence of the trend of increasing $\nli$ with increasing
$\feh$ has not been verified, we have decided to proceed with caution
in interpreting the data.  When determining if the models
are a good fit to the plateau star observations, we have removed the
$\feh$ trend from the data by correcting all stars to  $\feh =
-3.0$ before fitting our \li destruction isochrones to the
observations in the $\teffa$ plane.  The isochrones are fit to the data
using a $\chi^2$ technique, where the only free parameter is the
initial \li abundance.  Due to the real dispersion in the
data, it is impossible to obtain a good fit with any low order
function and so we have multiplied the error bars given by Thorburn by
1.3. When the error bars are increased by this amount, then the
straight line fit given by Thorburn
yields a reduced $\chi^2$ of 0.917, which is a
acceptable fit.  In determining the initial \li abundance implied by the
models, we fit the \li destruction isochrones to the raw data, using
the true error bars.  In performing our fits to the plateau data,
we ignore the 3 highly depleted plateau stars which were found by Thorburn.
The importance of the suggested
 positive correlation between \li abundance and metallicity and of the
highly depleted plateau stars will be
discussed in \S \ref{sect6p5}

\section{Model Construction \& Surface Boundary Conditions
\label{sectnew}}
In order to construct metal poor \li destruction isochrones, stellar models
with masses ranging from $0.55$ to
$0.80~M_\odot$ (in $\sim 0.03~M_\odot$ increments) and metallicities of
$Z = 10^{-4}$ and $10^{-5}$ were evolved from the pre-main sequence to  an
age of 18 Gyr, or main sequence turnoff.  Models which include no
mixing in the radiative region (standard), models which include
microscopic diffusion, and models which include both diffusion and
rotational mixing (combined) were calculated.  The prescription for
mixing used in the models is fully described in Chaboyer (1993) and
CDP, here we give a brief description.
The microscopic diffusion coefficients are from Michaud \&
Proffitt (1993).  The rotational mixing coefficients are taken to be
the product of a velocity estimate and the radius (Zahn 1992).  The
velocity estimates are similar to those used by Pinsonneault \ea (1989).
The rotating models include a general parameterization for the loss of
angular momentum at the surface due to magnetic stellar winds.  The
formulation is similar to that given by Kawaler (1988), except that a
saturation level, $\omega_{\rm crit}$ has been introduced.  If the
surface angular velocity is below $\omega_{\rm crit}$ then the rate
of angular momentum loss is that given by Kawaler (1988).
Above $\omega_{\rm crit}$, angular momentum loss occurs at a reduced
level compared to that given by Kawaler (1988).

The physics in the models are identical to those used in Chaboyer
(1993) and CDP.
Specifically, the energy producing nuclear reaction rates are from
Bahcall \& Pinsonneault (1992);  reaction rates for the $^6$Li, \li
and \be ~are from Caughlan \& Fowler (1988); the high
temperature opacities from Iglesias \& Rogers (1991); the low
temperature opacities (below $10^4$ K) are from Kurucz (1991); while
the  surface boundary conditions are determined using Kurucz model
atmospheres (Kurucz 1992; Howard, 1993).  The fit
between the stellar model and the atmosphere is made at an optical
depth of $\tau = 2/3$.
For temperatures above $10^6$ K, a relativistic degenerate, fully
ionized equation of state is use.  Below $10^6$ K, the single
ionization of $^1$H, the first ionization of the metals and both
ionizations of $^4$He are taken into account via the Saha equation.
It was found by Chaboyer (1993) that these models gave an excellent
description of the pre-main sequence \li depletion in solar metallicity stars.
Specifically, the \li depletion isochrones were in good agreement with
the Pleiades \li data of Soderblom \ea (1993) over the entire effective
temperature range investigated ($6500 \le \teffa \le 4000$ K).

In comparing to halo star \li observations, we must
pick an age for the stars.  Unfortunately, it is impossible to determine
ages of isolated field stars.   We have found that globular clusters
have ages ranging from 10 to 17 Gyr (Chaboyer, Sarajedini \& Demarque
1992).  It is likely that the metal poor halo stars and globular
clusters formed at approximately the same time.  If the hottest stars
(in a given metallicity range) for which \li observations have been
obtained, are interpreted as being near the main sequence turn off,
then a comparison to theoretical isochrones suggest an age of $\sim
18$ Gyr. The process of galaxy formation is not understood well enough
to give us a definitive age for the halo stars. For this reason, we
construct \li isochrones for the ages 10, 14 and 18 Gyr.

The models were fit to the data by varying the initial
\li abundance until the best match to the observations was obtained.
It was immediately clear that our models did not match the observed
depletion in the cool stars ($\teffa < 5300$ K).  Virtually no
depletion of \li occurred in the standard models, or diffusion
models.  This is very surprising, as previous work had a
significant depletion of \li for low mass stars (Deliyannis \&
Demarque 1991a,b).  The large amounts of \li depletion observed
in cool stars (and found by previous models) is due to the high
temperatures and densities achieved at the base of the convection zone
(particularly during the pre-main sequence) in low mass stars which
lead to substantial \li destruction.  Although some \li depletion
occurred in the cool star models which included rotation and diffusion,
it was at a level similar to the depletion suffered by the plateau
stars.  This problem with the models is  shown in Figure
\ref{haloatm}, where both standard models, and  combined models are
compared to the observations.
\begin{figure}[t]
\vspace*{8cm}
\caption{Comparison of standard and combined $^7$Li destruction
isochrones to the observations. The isochrones shown have an age of
18 Gyr.  The initial \li abundance was taken to be 2.25 (standard) and
3.0 (combined).}
\label{haloatm}
\end{figure}

To understand the source of this discrepancy, models were run with
different opacities, model atmospheres and  mixing length.  Changing
{}from the OPAL opacities (Iglesias \& Rogers 1991) to the LAOL
opacities (Huebner \ea 1977) had very little effect on the models.
Similarly, changing the low temperature opacities from Kurucz (1992)
molecular opacities to Cox \& Stewart (1970) opacities has only
a minor effect on the models and their \li depletion. Changing the
mixing length
within the stellar model (not the atmosphere)
to 3.0 did not substantially alter the \li depletion.
It is important to realize that these statements are not true in
general -- usually the \li depletion in cool stellar models is a sensitive
function of the assumed physics.  The reason that this is not true in
this particular case, is that the temperature at the base of the
convection zone in these models is far lower than that which is
required to deplete \Li.

Large amounts of overshoot at the base of the surface convection zone
lead to substantial \li depletion on the pre-main sequence.
We found that stellar models with an overshoot layer of $\sim 0.3$
pressure scale heights gave a reasonable match to the halo star \li
observations.  However, such a large overshoot layer is
clearly ruled out by the  observed \li depletion pattern in the Pleiades,
whose stars have just arrived on the main sequence. Good agreement
with the Pleiades \li observations occurs when the overshoot layer
has a depth of ${\rm H_p} \la 0.05$ pressure scale heights (CDP).
For this reason, we do not consider large amounts of overshoot as an
acceptable explanation of the discrepancy between the models and
observations.

Using a grey atmosphere (as opposed to a Kurucz atmosphere) for the
surface boundary conditions had a profound effect on the cool models
\footnote{In doing this comparison, the mixing length of the models
was kept fixed. This is a reasonable approximation, as a solar model
which is evolved with a grey atmosphere but with the calibrated
parameters from a Kurucz atmosphere run matches the observed solar
radius and luminosity to within 1\%.}.
The grey atmospheres were calculated in the standard manner
(Mihalas 1978).  We note that grey atmospheres
ignore the effects of convection.  The Kurucz atmospheres treat
convection using the mixing length approximation (with a mixing length
of 1.25).
The impact of the different model atmospheres on the temperature and
density at the base of the convection zone in a $M = 0.55~M_\odot$
stellar model (main sequence effective temperature $\sim 4800$ K).
is shown in Figure \ref{fig2}.  The models with Kurucz model
atmospheres have somewhat thinner convection zones, which leads to
lower temperatures and densities at the base of the convection zone.
This in turn leads to substantially less burning of \Li. Observations
indicate that this star should deplete \li by $> 2.2$ dex as
compared to the plateau stars. The model which uses the grey
atmosphere depletes \li by 2.3 dex, while the Kurucz model only
depletes \li by 0.25 dex.
We see that the amount of \li depletion in cool halo stars is
extremely sensitive to the surface boundary conditions.
\begin{figure}[t]
\vspace*{8cm}
\caption{Temperature and density at the base of the convection
zone for a $M=0.55~M_\odot$, $Z=10^{-4}$ stellar model.  Observations
indicate that this star should deplete \li by more than two orders of
magnitude.  The model which uses
Kurucz (1992) model atmospheres to determine the surface boundary
condition has convection zone base which is substantially cooler and
less dense than the model which uses a grey atmosphere.  The Kurucz
model depletes $\sim 0.25$ dex of \Li, while the grey atmosphere model
depletes the surface \li by $\sim 2.3$ dex.}
\label{fig2}
\end{figure}

It is clear that models constructed using
the grey atmosphere approximation are a better representation of
the halo star \li data then the Kurucz atmosphere models.  Thus,
we conclude that there could be
a problem with the surface boundary conditions
derived from metal-poor Kurucz atmospheres.
We caution, however, that no direct observations of ZAMS Pop II \li
abundances exist.  Thus, our models cannot be used to conclusively
claim that the surface boundary conditions
derived from metal-poor Kurucz atmospheres are in error.  For example,
there might exist a main sequence mixing mechanism which is more
efficient in cooler stars.   Such a mechanism could cause the models
with Kurucz atmospheres to match the present \li
observations in halo stars (C. Proffitt, private communication).
We note however, that no physical justification exists
for such a \li destruction mechanism, and such a \li destruction
mechanism does not exist in Pop I stars.
For this reason, we have
decided not to use the Kurucz model atmospheres in the remainder of this paper.
Henceforth, all of the models presented have been evolved using a grey
atmosphere,
in contrast to the models presented by CDP, which used the
Kurucz atmospheres.  We note that the good fit which CDP found to the
Pleiades \li observations is only true when the Kurucz atmospheres are
used -- the use of grey atmospheres leads to too much \li depletion as
compared to the Pleiades stars with $\teffa < 5500$ K.
It is also important to note that the difference in \li
depletion between the Kurucz and grey atmosphere only matters for
cool stars.
The \li depletion in the plateau stars are relatively
insensitive to the particular choice of surface boundary conditions.
Thus, the conclusions we draw in this paper regarding the \li
depletion in the plateau stars do not depend on our
choice of surface boundary conditions.

\section{Standard Models \label{sect6p2}}
$^7$Li destruction isochrones were constructed from stellar models which
incorporated various types and amounts of mixing in the radiative
region of the models.  Specifically, we evolved a set of standard
models (no mixing), two sets of models which included differing
amounts of diffusion, and two sets of combined models which included
differing amounts of rotational mixing and diffusion (combined models).
The mixing parameters used in the stellar models presented in this
paper are shown in Table \ref{tabsolar}.   The different sets of
models are referenced by a two letter code given in column 1 (the same
code as was used to identify the models in CDP).  Here, we comment
on the \li depletion in the standard models.
\begin{table}[t]
\begin{center}
\begin{tabular}{lcccccccc}
\multicolumn{9}{c}{TABLE 1}\\
\multicolumn{9}{c}{Stellar Model Parameters}\\
%\multicolumn{9}{c}{~~}\\
\tableline\tableline
& & & & & \multicolumn{4}{c}{Rotation Parameters\tablenotemark{b}}\\
\cline{6-9}
&\multicolumn{1}{c}{Mixing}&
\multicolumn{1}{c}{Overshoot}& & & & & &
\multicolumn{1}{c}{$\omega_{\rm crit}$\tablenotemark{f}}\\
\multicolumn{1}{c}{Model}&
\multicolumn{1}{c}{Length}&
\multicolumn{1}{c}{(H$_p$)}&
\multicolumn{1}{c}{$f_{\rm m}$\tablenotemark{a}}&
\multicolumn{1}{c}{Rotating?}&
\multicolumn{1}{c}{N\tablenotemark{c}}&
\multicolumn{1}{c}{$f_{\rm GSF}$\tablenotemark{d}}&
\multicolumn{1}{c}{$f_\mu$\tablenotemark{e}}&
\multicolumn{1}{c}{($10^{-5}s^{-1}$)}\\
\tableline
UU &1.83825&0.05 & 1.0 & Yes & 2.0 & 10  & 0.10 & 3.0\\
VN &1.80600&0.02 & 0.8 & Yes & 1.5 & 1.0 & 0.01 & 1.5\\
KB &1.86200&0.02 & 1.0 & No  & \rule[3pt]{0.4cm}{0.5pt}&
\rule[3pt]{0.4cm}{0.5pt}
&\rule[3pt]{0.4cm}{0.5pt}&\rule[3pt]{0.4cm}{0.5pt}\\
ND &1.83493&0.02 & 0.8 & No  &\rule[3pt]{0.4cm}{0.5pt}
&\rule[3pt]{0.4cm}{0.5pt}
&\rule[3pt]{0.4cm}{0.5pt}&\rule[3pt]{0.4cm}{0.5pt}\\
LA &1.72800&0.02 & 0.0 & No & \rule[3pt]{0.4cm}{0.5pt}
&\rule[3pt]{0.4cm}{0.5pt}
&\rule[3pt]{0.4cm}{0.5pt}&\rule[3pt]{0.4cm}{0.5pt}\\
\tableline\tableline
\end{tabular}
\end{center}

\tablenotetext{a}{Constant which multiplies the microscopic diffusion
coefficients of Michaud \& Proffitt 1993.}
\vspace*{-10pt}
\tablenotetext{b}{For a full description of the rotation parameters,
see CDP.}
\vspace*{-10pt}
\tablenotetext{c}{Power law index for the wind loss low. }

\tablenotetext{d}{Constant which multiplies the estimate for the GSF
diffusion coefficient.}

\tablenotetext{e}{Constant which determines the efficiency of
rotational mixing in the presence of mean molecular weight gradients.
Small values of $f_\mu$ imply efficient mixing.}

\tablenotetext{f}{Saturation level in the angular momentum loss law.}

\tablenum{1}

\label{tabsolar}
\end{table}

The low mass standard models deplete \li on the main sequence.
This results in more \li depletion in older isochrones, as is
demonstrated in Figure
\ref{haloage}.  Although the cool star fits are not perfect, it is
clear that the older isochrones produce a better match to the
observations.  This is not unexpected, as is unlikely that the
extremely metal poor stars shown here have an age of 10 Gyr.  An
isochrone with an age of $\sim 22$ Gyr would be a nearly ideal match
to the observations.  If the hottest stars (in a given
metallicity range) are interpreted as being near the main sequence
turn off, then a comparison to theoretical isochrones suggest an age
of $\sim 18$ Gyr.   Thus, it appears that the
standard models do not produce the \li depletion observed in cool
stars.  This might be due to uncertainties in the surface boundary
conditions.

The dependence of the amount of \li depletion on metallicity
is shown in Figure \ref{halomet}.  The $Z=10^{-5}$ isochrone ($\feh =
-3.28$) is completely flat between 6500 K and 5400 K.  This is at
variance with the observations.  In addition, the $Z=10^{-5}$
is considerably higher  than the $Z = 10^{-4}$ isochrone below 5600 K.
{}From the model standpoint, this is to be expected, because as we go to
lower metallicities the convection zones become shallower and it is
expected that less \li burning will occur.  However the few cool star
observations that are available do not show such a dramatic trend with
metallicity.  It is interesting to note that Deliyannis \& Demarque
(1991b) found the opposite trend with metallicity.  Their metal poor
models depleted {\rm more} \li than their metal rich models due to a
nonlinear dependence of the depth of the surface convection zone on
metallicity.  It is unclear why their models have this property.  We
are using identical \li nuclear  reaction rates.  The major difference
between this work and Deliyannis \& Demarque is that we are using the
OPAL (Iglesias \& Rogers 1991) and Kurucz (1992) opacities, while they
used the Cox \& Stewart (1970)  opacities.
\begin{figure}[p]
\vspace*{8cm}
\caption{Comparison of 10, 14 and 18 Gyr $^7$Li standard (LA)
isochrones with $Z = 10^{-4}$ to the observations. The initial \li
abundance was taken to be 2.23.}
\label{haloage}
\end{figure}
\begin{figure}[p]
\vspace*{8cm}
\caption{Comparison of $Z = 10^{-4}$ and $10^{-5}$ 18 Gyr
$^7$Li standard (LA) isochrones to the halo star observations. The
initial \li abundance was taken to be 2.25.}
\label{halomet}
\end{figure}

{}From Figures \ref{haloage} and \ref{halomet} we see that the best fit
to the data is obtained using the $Z = 10^{-4}$ 18 Gyr isochrone.
This isochrone still does a poor job of matching the cool stars, but
the fit to the plateau stars is quite acceptable with a  reduced
$\chi^2$ of 1.15, which is well within the $2\sigma$ range.  The
primordial abundance inferred from this isochrones is $\nli = 2.25\pm
0.012$.  Many of the other $Z = 10^{-4}$ isochrones produce acceptable
fits to the date, and yield primordial \li abundances  in the narrow
range $2.22 \le \nli \le 2.26$.  However, the $Z = 10^{-5}$ isochrones
are perfectly flat over the plateau temperature range.  A fit of these
isochrones to all of the plateau observations is rejected by the
$\chi^2$ analysis. This suggests the fact that the \li depletion in
our standard models have too great of a dependence on the metallicity.
If we restrict our attention to stars with $\feh \le -2.80$ (of which
there are only 21 stars), then the $\feh = -3.27$ are an acceptable fit.
In performing this fit to the extremely metal poor stars, we have
multiplied  the error bars by 1.1 (and  continue to correct for
the $\feh$ trend), to allow for the dispersion at a given $\teffa$.
The fit  is shown graphically in Figure \ref{halolowz} where we
compare the best fitting $Z = 10^{-5}$ isochrone to the extremely
metal poor plateau stars. It is clear from Figure \ref{halolowz}
that there are not enough observations of extremely metal poor stars
to rule out the possibility the flat isochrones produced by standard
models with $Z = 10^{-5}$.  However, it is likely that the the trend
of increasing $\nli$ with increasing $\teffa$ is true for these stars
as well.  It appears that the standards models do a poor job of
matching the observations.
\begin{figure}[t]
\vspace*{8cm}
\caption{A $\feh = -3.27$ \li standard (LA) destruction
isochrone is compared to
observations of $\nli$ in stars with $\feh \le -2.8$.  A $\chi^2$ fit
reveals that the isochrone is  acceptable at the $2\sigma$ level.
The initial \li abundance was taken to be 2.20.}
\label{halolowz}
\end{figure}

\clearpage
\section{Pure Diffusion Models \label{sect6p3}}
Microscopic diffusion causes \li to settle out of the surface
convection zone.  The time scale for this settling proportional to the
mass of the surface convection zone.   The mass of the surface
convection zone is a function of mass and metallicity with the mass of
the surface convection zone decreasing for high mass and/or low
metallicity stars.  Thus, uninhibited microscopic diffusion will lead to
a downward curvature of the \li destruction isochrones at the hot edge
of the plateau.  The observational database available prior to
Thorburn (1994) did not rule out such a trend.  As such, previous
studies of the effects of microscopic diffusion on halo star \li
abundances obtained satisfactory agreement with the observations
(Proffitt  \& Michaud 1991; Chaboyer \ea 1992).

The new data rule out any trend of decreasing $\nli$ with
increasing $\teffa$ in the plateau temperature range ($\teffa > 5590$
K).  This is shown graphically in Figure \ref{halodiff} where we plot
our best fitting pure diffusion models to the observations.
\begin{figure}[t]
\vspace*{8cm}
\caption{18 Gyr
ND \li destruction isochrones are compared to the observations.  The
diffusion coefficients have been multiplied by 0.8, but the isochrones
still have too much curvature and are rejected by a $\chi^2$ analysis.
The initial \li abundance as determined by the $\chi^2$ analysis was
2.40 for $Z = 10^{-4}$ and 2.45 for $Z = 10^{-5}$.}
\label{halodiff}
\end{figure}
Clearly,
these isochrones do not reproduce the mean trend visible in the data.
A $\chi^2$ fit of the diffusive isochrones to the plateau data rejects all of
the isochrones at a very high probability.  The results of the
$\chi^2$ analysis are presented in Table \ref{tabhalodiff} from which
it is clear that diffusive isochrones do not match the observations.
We have also performed a $\chi^2$ analysis of the diffusive isochrones
which use the Kurucz (1992) model atmospheres as surface boundary
conditions, and find that they too are rejected as good fits to the
$\nli$ data with $\teffa > 5590$ K.
We conclude that the diffusion coefficients given by Michaud \& Proffitt
(1993) yield too much depletion of \li for the hot stars.
The models shown in Figure \ref{halodiff} have had their diffusion
coefficients artificially lowered by 25\%, which demonstrates that
even an error of $\sim 25\%$ in the diffusion coefficients does not
effect our conclusion that diffusion leads to an over depletion of
\li in the hot stars.
\begin{table}[t]
\begin{center}
\begin{minipage}{14.3cm}
\begin{tabular}{ccccccc}
\multicolumn{7}{c}{TABLE 2} \\
\multicolumn{7}{c}{$\chi^2$ Fits of the Diffusive Isochrones to
Plateau Stars}\\
\tableline\tableline
\multicolumn{3}{c}{Isochrones}&
\multicolumn{1}{c}{Observational}&
\multicolumn{1}{c}{Reduced}& &
\multicolumn{1}{c}{Initial \li}\\
\cline{1-3}
\multicolumn{1}{c}{Age (Gyr)}&
\multicolumn{1}{c}{$\feh$}&
\multicolumn{1}{c}{$f_{\rm m}$\tablenotemark{a}}&
\multicolumn{1}{c}{Data Set}&
\multicolumn{1}{c}{$\chi^2$}&
\multicolumn{1}{c}{Probability\tablenotemark{a}}&
\multicolumn{1}{c}{Abundance}\\
\tableline
10 & $-2.28$ & 1.0& all stars & 1.86 & $<0.001$ & 2.36\\[0.5pt]
14 & $-2.28$ & 1.0& all stars & 1.84 & $<0.001$ & 2.40\\[0.5pt]
18 & $-2.28$ & 1.0& all stars & 1.57 & $<0.01$  & 2.44\\[0.5pt]
10 & $-2.28$ & 0.8& all stars & 1.75 & $<0.001$ & 2.34\\[0.5pt]
14 & $-2.28$ & 0.8& all stars & 1.73 & $<0.001$ & 2.37\\[0.5pt]
18 & $-2.28$ & 0.8& all stars & 1.49 & $<0.01$  & 2.40\\[0.5pt]
14 & $-2.28$ & 0.8&$\feh >-2.8$& 1.86 & $<0.001$ & 2.37\\[0.5pt]
18 & $-2.28$ & 0.8&$\feh >-2.8$& 1.57 & $<0.001$ & 2.41\\[0.5pt]
14 & $-3.28$ & 0.8& all stars & 2.59 & $<0.001$ & 2.42\\[0.5pt]
18 & $-3.28$ & 0.8& all stars & 1.64 & $<0.001$ & 2.45\\[0.5pt]
10 & $-3.28$ & 0.8&$\feh <-2.8$& 2.12 & $<0.01$ & 2.39\\[0.5pt]
14 & $-3.28$ & 0.8&$\feh <-2.8$& 2.32 & $<0.001$ & 2.44\\[0.5pt]
18 & $-3.28$ & 0.8&$\feh <-2.8$& 2.31 & $<0.001$ & 2.46\\[0.5pt]
\tableline\tableline
\end{tabular}
\end{minipage}
\end{center}

\tablenotetext{a}{Constant which multiplies the microscopic diffusion
coefficients of Michaud \& Proffitt 1993.}
\vspace*{-10pt}
\tablenotetext{b}{Probability of a random data
set exceeding the reduced $\chi^2$ when compared to the isochrone.}

\tablenum{2}
%\caption[$\chi^2$ Analysis of the Diffusive Isochrones]{}
\label{tabhalodiff}
\end{table}

\section{Combined Rotation and Diffusion Models \label{sect6p4}}
Rotational mixing in the models can lead to the inhibition of
microscopic diffusion near the surface.  Thus, it is likely that
the \li depletion isochrones for the combined models will not possess
the strong downward curvature with increasing $\teffa$ that is observed
in pure diffusion models.  Previous studies of the impact of
rotational mixing (without diffusion)
on  halo stars \li abundances found a rather large, uniform
depletion (Pinsonneault \ea 1992) and so it is important to study the
combined effect of both mixing processes. Chaboyer (1993)
 determined that the model VN parameters (see Table
\ref{tabsolar}) provided the best match to the \li observations in
the Sun and young  cluster stars.  For this
reason, we have chosen to evolve a set of low metallicity models
using the same VN parameters (except for the use of a grey
atmosphere).  The initial rotational velocity of the models is a free
parameters, we have evolved models with $V_{rot} = 10,~30$ and 50
km/s, which encompasses the range observed in T-Tauri stars (Bouvier
\ea 1993).  The $Z = 10^{-4}$ isochrones provide a good match to the
observations and are shown in Figure \ref{haloVN}.
\begin{figure}
\vspace*{8cm}
\caption{Model VN isochrones with ages of 14 and 18 Gyr
are compared to the
observations.  For each age, two isochrones are shown, corresponding
to initial rotation velocities of 10 and 30 km/s on the pre-main
sequence.  The 30 km/s isochrones are $\sim 0.3$ dex below the 10 km/s
isochrones for $\teffa > 5590$ K.
The initial \li abundance determined by a $\chi^2$ fit to
the plateau stars is 3.03 for the 14 Gyr, 10 km/s isochrone and
3.13 for the 18 Gyr, 10 km/s isochrone.}
\label{haloVN}
\end{figure}
The 14 Gyr
observations show a slight downward trend at the hot edge of the
plateau which is not present in the 18 Gyr isochrones.  The  $\chi^2$
fit to the plateau stars  reveals that the 14 and 18 Gyr
VN isochrones are a good match to the observations.  The initial
\li abundance  required to match the observations is $\nli = 3.03$ and
$\nli = 3.13$ for the 14 and 18 Gyr isochrones respectively.  These
values are similar to those found by Pinsonneault \ea (1992) for pure
rotation models.  In contrast to the standard models presented in \S
\ref{sect6p2}, these combined isochrones do a good job of fitting the
observations of \li in the cool stars ($\teffa < 5500$ K).

An examination of the time and mass dependence of the
\li depletion reveals the reason for the flattening of the isochrones
at 18 Gyr.  The \li depletion as a function of age is plotted in
Figure \ref{haloliage} for stars with masses of $0.75$, $0.72$ and
$0.70~M_\odot$.
\begin{figure}
\vspace*{8cm}
\caption{$^7$Li depletion as a function of age for
$Z = 10^{-4}$ combined models.}
\label{haloliage}
\end{figure}
These stars lie on the hot half of the plateau and so
are responsible for the flattening of the isochrones.
During the early pre-main sequence ($t < 10$ Myr) all
models experience a modest amount of surface \li depletion.   Low mass
stars deplete somewhat more \Li.    Starting near the ZAMS, ($t \sim 60$
 Myr) rotational mixing depletes the
surface \li abundance.  This depletion is more efficient in the lower
mass stars. The lower mass star  have deeper convection zones, hence
the \li needs to be transported a smaller distance  before being destroyed.
  After $\sim 6$ Gyr the rotational mixing time scale is very long and
the depletion due to diffusion becomes important.
The diffusive \li depletion time scale is proportional to the mass of
the surface convection zone. The convective envelope mass (and hence,
the time scale for \li depletion) declines on the main sequence.
A $Z = 10^{-4}$, $M = 0.75~M_\odot$ stellar model has a
convective envelope mass less than $4\times 10^{-3}~M_\odot$ at 6 Gyr.
At this point the mixing due to diffusion causes the
\li surface depletion rate to be significantly higher in a $0.75~M_\odot$ model
than a $0.72~M_\odot$ model. At 11 Gyr, the surface \li abundance is
very similar in these two models.  By 14 Gyr, the surface \li abundance of
the $0.75~M_\odot$ model is well below that of the $0.72~M_\odot$
model, and so the \li destruction isochrone shows a downward curvature
at 14 Gyr.  The $0.75~M_\odot$ model has a turnoff age of $\sim 15$
Gyr, and so does not contribute to our 18 Gyr \li destruction
isochrone.  Instead, the hot edge of the 18 Gyr isochrone is located
by the $0.72~M_\odot$ model whose convection zone depth is
considerable larger than the $0.75~M_\odot$ model.  The
diffusion time scale is such that by an age of 18 Gyr, the total \li
depletion in a $0.72~M_\odot$ model is similar to a $0.70~M_\odot$ model
leading to a \li destruction isochrone which does not turn down at hot
temperatures.

Over the plateau star $\teffa$ range ($\teffa \ga 5600$ K),
the $V_i = 30$ km/s isochrones
are depleted $\sim 0.3$ dex more than the 10 km/s isochrones.   Since
stars are likely to have a range of initial rotation velocities, this
implies that the combined models predict a dispersion in the observed
\li abundance.  It is impossible to determine the initial rotation
velocity distribution for halo stars.  Observations of rotation
periods in low mass T-Tauri stars indicate that 2 out of 19 single
stars (10\%) have rotation velocities greater than or equal to 30 km/s
(Bouvier 1991). No stars were found to rotate slower than 10 km/s
(but there is an observational bias against  long periods)
while 11 stars (58\%) had rotation velocities  in the range 10 -- 20
km/s.  We have run a few models with $V_i = 20$ km/s which indicate
that these stars experience $\sim 0.2$ dex more depletion than the 10
km/s models. Assuming that the halo stars have a similar
distribution in initial rotation velocities as the T-Tauri stars, the
above numbers suggest that our rotation models should have a
dispersion of $\sim 0.15$ dex, which is similar to the 0.11 dex
dispersion present in the observations before the metallicity trend
is removed.  However, after the metallicity trend is removed,
the dispersion in the observations is 0.07 dex and it would appear
that the models predict too large of a dispersion.  A more careful
study of predicted dispersion is required before making any definitive
conclusions.

The isochrones change dramatically for the $Z = 10^{-5}$ models.
At a given $\teffa$, the convection zone depths in these models are
shallower than the $Z = 10^{-4}$ models.  Hence, microscopic diffusion
plays a more important role in these models.  This is shown in Figure
\ref{haloVNlowz} where the 10, 14 and 18 Gyr $Z = 10^{-5}$ isochrones
are plotted.
\begin{figure}
\vspace*{8cm}
\caption{Model VN isochrones with ages of 10, 14 and 18 Gyr
and $Z = 10^{-5}$ are compared to the
observations.  The $V_i = 10$ km/s isochrones are shown.
The 30 km/s isochrones are $\sim 0.3$ dex below the 10 km/s
isochrones for $\teffa > 5590$ K.
The initial \li abundance was taken to be 2.87.}
\label{haloVNlowz}
\end{figure}
The \li depletion in these plateau temperature range is
approximately 0.1 dex {\it less than} that found in the $Z = 10^{-4}$
isochrones.  This is in the opposite direction to the $\nli$ -- $\feh$
trend observed in the data.  In addition, there is a large downward
trend present in all of these isochrones which is not present in the
observations.  A $\chi^2$ analysis rejects these fits at the greater
than $3\sigma$ level.  A fit to those  stars with
$\feh \le -2.8$ also rejects these isochrones at a greater than
$3\sigma$ level.  This is graphically illustrated by Figure
\ref{halolowzVN} where the metal poor isochrones are compared to the
 stars which have $\feh \le -2.8$ and $\teffa > 5700$ K.
This clearly demonstrates that the only parameters which gave
a good representation of the \li depletion pattern observed in young
clusters fail to match the extremely metal poor halo star observations.
\begin{figure}
\vspace*{8cm}
\caption{Model VN isochrones with ages of 10, and 14  Gyr
and $\feh = -3.28$ are compared to
observations of stars which have $\feh \le -2.8$ and $\teffa >5700$ K.
The $V_i = 10$ km/s isochrones are shown.
The initial \li abundance was taken to be 2.87.}
\label{halolowzVN}
\end{figure}

We have also constructed halo star models using model UU parameters
{}from Table \ref{tabsolar} in order to test the sensitivity of our
conclusions to the rotational mixing parameters.
  The rotational mixing is much more
sensitive to gradients in the mean molecular weight in these models as
compared to the VN models presented previously.  The $Z = 10^{-4}$
model UU isochrones are shown in Figure \ref{haloU}.
\begin{figure}[t]
\vspace*{8cm}
\caption{Model UU isochrones with ages of 10, 14 and 18 Gyr
and $Z = 10^{-4}$  are compared to the
observations.  The $V_i = 10$ km/s isochrones are shown.
The 30 km/s isochrones are $\sim 0.25$ dex below the 10 km/s
isochrones for $\teffa > 5590$ K.
The initial \li abundance was taken to be 3.02}
\label{haloU}
\end{figure}
Since diffusion
of $^4$He out of the surface convection zone leads to a large mean
molecular weight gradient at the base of the convection zone, the
rotational mixing is overwhelmed by the diffusion at a much earlier
age than in the VN models.  This is clear from the isochrones in
Figure \ref{haloU}, all of which have some downward curvature to them
over the plateau effective temperature range.  The $\chi^2$ analysis
reject the 10 and 14 Gyr isochrones as acceptable fits to the data.
The 18 Gyr isochrone has less  curvature and is an acceptable fit to
the observations at the $2\sigma$ level.  It is clear that the VN
isochrones are a better match to the data, which strengthens the CDP
conclusion  that the VN parameters provide the
best fit to the observations. The primordial \li abundance
determined by the fit is $\nli = 3.02$, which is similar to that
inferred from the 14 Gyr VN isochrone.

\clearpage
\section{Summary \label{sect6p5}}
It is clear that the  halo star \li observations of Thorburn (1994)
provide an excellent database with which to test stellar evolution
models.  Stellar evolution models which use the Kurucz (1992)
model atmospheres as the surface boundary condition deplete virtually
no \li over the entire temperature range studied ($6500 < \teffa < 4800$
K).  This is clearly at odds with the observations,
which show substantial \li depletion below $\teffa = 5600$ K.
We conducted extensive tests which determined that the models which
used grey atmospheres did not suffer from this problem.
Changing the opacities or mixing length did not
substantially alter the \li depletion pattern.  However,
models which use the grey atmosphere approximation as
the surface boundary condition provide a good match to the
observations. It appears that the low metallicity Kurucz (1992)
atmospheres  do not provide the appropriate boundary conditions for
our stellar evolution models.  This is in sharp contrast to the solar
metallicity models studied by CDP, where the use of
the Kurucz atmospheres, and not the grey atmospheres,
provided an excellent match to the observations.
We caution, however, that no direct observations of ZAMS Pop II \li
abundances exist.  Thus, our models cannot be used to conclusively
claim that the surface boundary conditions
derived from metal-poor Kurucz atmospheres are in error.

Standard models (with Kurucz or grey atmospheres)
and $\feh = -2.28$
do a good job of fitting the  observed \li abundances in stars
hotter than $\sim 5600$ K.  They
predict a primordial \li abundance of $\nli = 2.24\pm 0.03$.
The cooler stars have somewhat lower abundances than the isochrones
predict.  This is not too serious a problem since  the amount of
depletion in the low mass models is a strong function of the amount of
overshoot assumed at the base of the convection zone, and of the
atmospheres used to calculate the surface boundary conditions.  Of
greater concern is the fact that decreasing the metallicity in the
models leads to a decrease in the amount of \li depletion.  Although
there are not many stars with $\feh < -2.8$ it appears that the
observations do not possess this property.

The observations provide a very strict test of models which include
microscopic diffusion, for these models predict a downward curvature
in the \li destruction isochrones at hot temperatures,
regardless of the choice of choice of atmospheres.
Even allowing for a dispersion in the initial \li abundance,
the observations clearly rule out models which include uninhibited
microscopic diffusion of \li from the surface of the star.
This has been believed to have important implications for the
diffusion of $^4$He, for the time scale for the diffusion is very similar
for $^4$He and \li (Chaboyer \ea 1992). However, the situation is
different when rotation is considered.  The
rotational mixing
in our models causes an order of magnitude depletion in \Li, while
diffusion at the hot end of the plateau depletes \li by a factor of
$\sim 2$. Thus, the large overall depletion of \li caused by rotational
mixing over the entire effective temperature range of the \li plateau
tends to overwhelm the \li depletion caused by diffusion in the hot
plateau stars.  By contrast, $^4$He is not depleted by rotational
mixing, and so it is difficult to make statements regarding the
depletion of $^4$He, based on the constraints from \li observations.
For example, an $0.72~{\rm M}_\odot$ stellar model, at an age of 16
Gyr has a surface $^4$He mass fraction  0.230 (standard, model LA);
0.141 (pure diffusion, model ND) and 0.154 (combined, model VN).
The effective temperature of this model is 6292 K (standard),
6280 K (diffusion) and 6288 K (combined).  Thus, we see that
inhibition of diffusion by rotational mixing in hot halo stars is a
relatively minor effect, reducing the surface depletion of $^4$He by
$\sim 10\%$.  However, the shape of the \li depletion isochrones are
substantial altered by the presence of rotational mixing, leading to
good agreement with the observations.

The diffusion of
$^4$He from the surface of the star has a dramatic effect on effective
temperature of the stellar evolution model.  In a previous paper, we
found that uninhibited diffusion of $^4$He from the surface altered the
shape of our isochrones (in the $M_V$, ${\rm B - V}$ plane) such that
fits to globular cluster colour magnitude diagrams
suggested an age reduction of 2 -- 3 Gyr
compared to standard isochrones (Chaboyer \ea 1992).  However, the age
reduction derived using the \dv technique (which relies on the
luminosity of the models, {\it not the effective temperature\/}) was
very small, $\sim 0.5$ Gyr.
In order to see if these conclusions are altered by the present
calculations, we have calculated isochrones, in a similar manner to
Chaboyer \ea (1992).  Sample isochrones, for an age of 16 Gyr, are
shown in Figure \ref{iso}.  We see that the introduction of rotational
mixing does not substantial alter the shape of the isochrone (ie --
the pure diffusion isochrone is quite similar to the isochrone which
includes rotation and diffusion). We find the following turnoff
magnitudes and colours for the 16 Gyr isochrone: ${\rm M_v} = 3.78$,
${\rm B - V} = 0.341$ (standard); ${\rm M_v} = 3.86$,
${\rm B - V} = 0.362$ (diffusion); and ${\rm M_v} = 3.85$,
${\rm B - V} = 0.360$ (combined).  Hence, these calculations do not
alter the conclusions of Chaboyer \ea (1992) regarding the effect of
diffusion on age estimates of globular clusters.   Figure \ref{iso}
\begin{figure}[t]
\vspace*{8cm}
\caption{Comparison of 16 Gyr isochrones, which were
calculated from stellar models with different mixing prescriptions.
Shown are standard (model LA, dash-dot line),
pure diffusion (model ND, dashed line) and combined,
rotation and diffusion (model VN, solid line) isochrones.  The
combined model is quite similar to the pure diffusion model.}
\label{iso}
\end{figure}
demonstrates that even in the presence of rotation, the dominant
effect of diffusion is on the colours of the isochrones.  Hence, age
determination techniques which rely on using colours of the isochrones
(such as isochrone fitting) imply that diffusion lowers the age
estimate of globular clusters by 2 -- 3 Gyr.  In contrast, if we were
to date globular clusters using the main sequence turn off luminosity,
and an independent distance estimate, we would find an age reduction
of 1 Gyr.  The age reduction derived using the \dv technique
is 0.5 Gyr, for this age determination relies
on calculating the horizontal branch luminosity,
which is also slightly lowered  by diffusion.

Rotational mixing can inhibit the microscopic diffusion leading to
less curvature in the \li depletion isochrones for $\teffa > 5600$ K.
The $\feh = -2.28$ VN stellar models which include both diffusion and
rotational mixing provide an excellent match to the mean trend in
$\teffa$ which is present in the observations.
Both the plateau stars and the heavily depleted cool stars are well
fit by these models.  The dispersion at a give $\teffa$ predicted by
the models due to varying initial rotation velocities is roughly
consistent with what is observed.  The rotational mixing leads to
considerable \li depletion in these models and the primordial \li
abundance inferred from these models is $\nli = 3.08$, with an
uncertainty of $\sim 0.1$ dex due to uncertainties in the rotational
mixing coefficients.  This is similar to the value found by
Pinsonneault \ea (1992) for models which included rotational mixing,
but not microscopic diffusion.

However, the $\feh = -3.28$ isochrones reveal problems with
the combined models.  These isochrones
predict a trend of decreasing $\nli$ with increasing $\teffa$ which
is clearly not present in the observations.  This indicates that the
diffusion (which becomes more efficient at depleting \li in hot stars
as the metallicity decreases) is not being inhibited enough by the
rotational mixing.  Thus suggests that the rotational mixing is more
efficient (particularly in very low metallicity stars) than has been
modeled here.   Also troubling is the fact that models with low
metallicities deplete {\it less} \li than the more metal rich models,
for the observations of plateau stars reveal that the  \li abundance
{\it decreases} with lower metallicities.  The positive correlation
of $\nli$ with $\feh$ is not compatible with these models unless
\li production has occurred.

Pure diffusion models are ruled out by \li observations in young cluster
stars and the Sun (CDP) and, as was shown in this paper, by the
halo star \li observations.  The standard models provide a reasonable
fit to the halo star \li observations, but, as was shown by CDP,
fail to account for the \li depletion observed in the Sun, or young
cluster stars.  The combined models with rotational mixing and
diffusion are a good match \li observations in stars with
$\feh \ga -2.3$, but do not agree with the most metal poor halo star
observations.  It is clear that the observed \li abundances provide a
powerful test of stellar evolution models; a test which some of our
models do not pass. The reason for this breakdown of our lowest
metallicity combined models is
uncertain.  It could be that rotational mixing is more efficient in
metal poor stars than is indicated by our prescription.
Another possibility may be related to the surface boundary conditions
in the models.  The \li depletion in the models
can depend sensitively on the detailed atmospheric structure used in
the models. We note that the effects of convection, which is not well
understood, become increasingly important in low metallicity atmospheres.

In addition none of our models (standard, diffusive or combined rotation and
diffusion) predict a decrease in $\nli$ with decreasing $\feh$ which
is present in the observations (Thorburn 1994).    This
trend is compatible with the ideal that \li production has occurred due
to galactic cosmic ray spallation (Thorburn 1994). If this is true,
then the primordial \li abundance must be lower than that obtained
{}from fits to  all of the plateau stars.  If \li production has occurred,
then the observed dispersion in $\nli$ is easily explained by an age
range of $\sim 2$ Gyr in stars with the same metallicity. We caution
however, that the \li abundances were determined
using the Kurucz model atmospheres, which include the effects of
convection
using the mixing length approximation.  Convection becomes
increasingly vigorous in more metal-poor stars and a
 better understanding of
atmospheric models is required to answer the question of decreasing \li
abundance with decreasing metallicity.

The presence of three highly depleted `plateau' stars demonstrate that
substantial \li depletion is occurring in some plateau stars.  Standard
stellar evolution models do not predict any depletion in this
temperature range.  Young cluster observations
clearly demonstrate that \li destruction is occurring on the main
sequence at a rate much larger than that predicted by standard models.
This clearly points to the existence of an additional mixing mechanism
in the radiative zones of some stars.  If this mixing mechanism is
related to rotation, our prescription for rotational mixing fails to
match these observations. The understanding of mixing in the radiative
zones of stars remains one of the outstanding problems in stellar
evolution theory.

\acknowledgments
We are indebted  to Julie Thorburn, who gave us her Li observations
in advance of publication.  In addition, we would like to thank the
referee, Charles Proffitt for his thoughtful comments which improved
the presentation of this paper.
Research supported in part by NASA grants NAG5--1486, NAGW--2136
NAGW--2469 and NAGW-2531.

\parskip 0em
\small
%\vspace*{-20pt}


\begin{references}
\vspace*{-12pt}
\reference Bahcall, J.N., \& Pinsonneault, M.H. 1992, Reviews of Modern
Physics, 64, 885
\vspace*{-5pt}
\reference Bouvier, J. 1991, in Angular Momentum Evolution of Young Stars,
ed. S. Catalano \& J.R. Stauffer,
(Dordrecht: Kluwer), 41
\vspace*{-5pt}
\reference Bouvier, J., Cabrit, S., Fern\'{a}ndez, Mart\'{i}n, E.L.
and Matthews, J.M.  1993, A\&A, 272, 176
\vspace*{-5pt}
\reference Caughlan, G.R. \& Fowler, W.A. 1988, Atomic Data \& Nuclear
Data Tables, 40, 283
\vspace*{-5pt}
\reference Chaboyer, B.C. 1993, PhD thesis, Yale University
\vspace*{-5pt}
\reference Chaboyer, B.C., Deliyannis, C. P., Demarque, P.,
Pinsonneault, M. H., \& Sarajedini, A.  1992,  ApJ,
388, 372
\vspace*{-5pt}
\reference Chaboyer, B.C., Demarque, P. \& Pinsonneault, M. H 1994a,
submitted to ApJ (CDP)
\vspace*{-5pt}
\reference Chaboyer, B.C., Demarque, P. \& Pinsonneault, M. H 1994b,
submitted to ApJ (CDP)
\vspace*{-5pt}
\reference Chaboyer, B.C., Sarajedini, A., \& Demarque, P. 1992,
ApJ, 394, 515
\vspace*{-5pt}
\reference Cox, A.N., \& Stewart, J.N. 1970, ApJS, 31, 271
\vspace*{-5pt}
\reference Deliyannis, C.P., \& Demarque, P. 1991a, ApJ, 379, 216
\vspace*{-5pt}
\reference Deliyannis, C.P., \& Demarque, P.  1991b, ApJL, 370, L89
\vspace*{-5pt}
\reference Hobbs, L.M., Welty, P.E. \& Thorburn, J.A. 1991, ApJ, 373, L47
\vspace*{-5pt}
\reference Hobbs, L. M. \& Thorburn, J.A. 1991, AJ, 102, 1070
\vspace*{-5pt}
\reference Howard, J.M. 1993, PhD thesis, Yale University
\vspace*{-5pt}
\reference Huebner, W.F., Merts, A.L., Magee, N.H., \& Argo, M.F. 1977,
Los Alamos Opacity Library, Los Alamos Scientific Laboratory Report,
No. LA--6760--M
\vspace*{-5pt}
\reference Iglesias, C.A. \& Rogers, F.J. 1991, ApJ, 371, 408
\vspace*{-5pt}
\reference Kawaler, S.D. 1988, ApJ, 333, 236
\vspace*{-5pt}
\reference Krauss, L.M. \& Romanelli, P.  1990 ApJ, 358, 47
\vspace*{-5pt}
\reference Kurucz, R.L. 1991, in Stellar Atmospheres: Beyond Classical
Models, ed. L. Crivellari, I. Hubeny, D.G. Hummer, (Dordrecht: Kluwer) 440
\vspace*{-5pt}
\reference Kurucz, R.L. 1992, in IAU Symp. 149, The Stellar
Populations of Galaxies, ed. B. Barbuy, A. Renzini, (Dordrecht: Kluwer), 225
\vspace*{-5pt}
\reference Malaney, R.A. \& Mathews, M.N. 1993, Physics Reports, 229,
145
\vspace*{-5pt}
\reference Michaud, G. \& Proffitt, C.R. 1993, in Inside the Stars, IAU
Col. 137, ed. A. Baglin \& W.W. Weiss (San Francisco: ASP), 246
\vspace*{-5pt}
\reference Mihalas, D. 1978,Stellar Atmospheres, 2nd Edition, (San
Francisco:  W.H.~Freeman \& Co.)
\vspace*{-5pt}
\reference Pinsonneault, M.H., Deliyannis, C. P., \& Demarque, P. 1992, ApJS,
78, 181
\vspace*{-5pt}
\reference Pinsonneault, M.H., Kawaler, S.D., Sofia, S., \& Demarque, P. 1989,
ApJ, 338, 424
\vspace*{-5pt}
\reference Proffitt, C.R., \& Michaud, G. 1991, ApJ, 371, 584
\vspace*{-5pt}
\reference Rebolo, R., Molaro., P., \& Beckman, J. E. 1988, A\&A, 192, 192
\vspace*{-5pt}
\reference Smith, M.S., Kawano, L.H. \& Malaney, R.A. 1993, ApJS, 85, 219
\vspace*{-5pt}
\reference Smith, V.V., Lambert, D.L, \& Nissen, P.E. 1992, ApJ, 408, 262
\vspace*{-5pt}
\reference Soderblom, D.R., Jones, B.F., Balachandran, S., Stauffer, J.R.,
Duncan, D.K., Fedele, S.B. \& Hudon, J.D. 1993,  AJ, 106, 1059
\vspace*{-5pt}
\reference Spite, M., Molaro, P., Fran\c cois, P. \& Spite, F. 1993,
A\&A, 271, L1
\vspace*{-5pt}
\reference Spite, F. \& Spite, M. 1982, A\&A, 115, 357
\vspace*{-5pt}
\reference Spite, F. \& Spite, M.  1986, A\&A, 163, 140
\vspace*{-5pt}
\reference Thorburn, J.A. 1994, ApJ, 421, 318
\vspace*{-5pt}
\reference Thorburn, J.A., Hobbs, L.M, Deliyannis, C.P. \& Pinsonneault,
M.H. 1993, ApJ, 415, 150
\vspace*{-5pt}
\reference Walker, T.P., Steigman, G., Schramm, D.N., Olive, K.A.
\& Kang, H.S.  1991, ApJ, 376 51
\vspace*{-5pt}
\reference Witten, E. 1984, Phys. Rev. D, 30. 272
\vspace*{-5pt}
\reference Zahn, J.-P. 1992, to appear in  Astrophysical Fluid Dynamics,
Les Houches XLVII, ed. J.-P. Zahn \& J. Zinn-Justin (Elsevier Science Pub.)

\end{references}
\end{document}